\documentclass[]{spie}  

\usepackage{amsmath,amsfonts,amssymb}
\usepackage{graphicx}
\usepackage[colorlinks=true, allcolors=blue]{hyperref}

\title{SAGE: using CubeSats for Gravitational Wave Detection}

\author[a,b]{S. Lacour}
\author[a]{M. Nowak}
\author[c]{P. Bourget}
\author[a]{F. Vincent}
\author[c]{A. Kellerer}
\author[a]{V. Lapeyrere}
\author[a]{L. David}
\author[d]{A.~Le~Tiec}
\author[a]{O. Straub}
\author[c]{J. Woillez}

\affil[a]{LESIA, Observatoire de Paris, PSL Research University, CNRS, 5 place Jules Janssen, 92190 Meudon}
\affil[b]{Max Planck Institute for extraterrestrial Physics, Giessenbachstr. 1, D-85748 Garching, Germany}
\affil[c]{European Southern Observatory, Karl-Schwarzschild-Strasse 2, 85748 Garching, Germany}
\affil[d]{LUTH, Observatoire de Paris, PSL Research University, CNRS, 5 place Jules Janssen, 92190 Meudon}

\authorinfo{Contact E-mail: sylvestre.lacour@obspm.fr}

\begin{document} 
\maketitle

\begin{abstract}
SAGE (SagnAc interferometer for Gravitational wavE) is a fast track project for a space observatory based on multiple 12-U CubeSats in geostationary orbit. The objective of this project is to create a Sagnac interferometer with 73\,000\ km circular arms. The geometry of the interferometer makes it especially sensitive to circularly polarized gravitational waves at frequency close to 1Hz. The nature of the Sagnac measurement makes it almost insensitive to position error, allowing spacecrafts in ballistic trajectory. The light source and recombination units of the interferometer are based on compact fibered technologies, without the need of an optical bench. The main limitation would come from non-gravitational acceleration of the spacecraft. However, conditionally upon our ability to post-process the effect of solar wind, solar pressure and thermal expansion, we would detect gravitational waves with strains down to $10^{-21}$ over a few days of observation.
\end{abstract}


\section{INTRODUCTION}
\label{sec:intro}  

Gravitational-wave (GW) astronomy will be a major observing window on the Universe 
in the next decades. The ground-based detectors that obtained
the first direct detection of gravitational waves~\cite{abbott16} are sensitive around $10^3$~Hz,
while the future Laser Interferometer Space Antenna (LISA) 
will reach maximum sensitivity at $\approx 10^{-2}$~Hz.
The gap in between these two ranges is and has been the subject of investigation
by various groups, with quite a few space-interferometer proposals being published (for a review of gravitational wave detection in space, see reference~\cite{ni16}).

The traditional problem with space missions is the difficult science/cost ratio. Flagship missions like LISA can be funded thanks to the tremendous scientific return it will have on the community. Middle size/middle cost space missions can also exist, providing they give a consequent, but not risky, return on investment. However, low cost/small space missions have difficulties to exist since the science case is often limited, and the low cost implies increased risk. This is especially relevant in the context of GW detection. The space community will be focused for decades (and more) on the LISA mission, and it will be hard to find money for another medium-range mission.

In this context, we still want to try to open the gravitational window for waves around 1\,Hz. Attempts to go to lower frequencies with ground-based detectors are under study, but past the seismic noise, a ground detector will always be limited by Earth gravitational disturbances. Our solution to these difficulties is a low cost space mission. This is not a new concept: it has already been proposed to use off-the-shelf satellites \cite{tinto15}. However, with SAGE, we propose to go several steps further in simplification. The main simplifying step is the use of the CubeSat standard. But it is only possible because other steps are taken:

\begin{figure}
\centering
\includegraphics[width=0.5\textwidth]{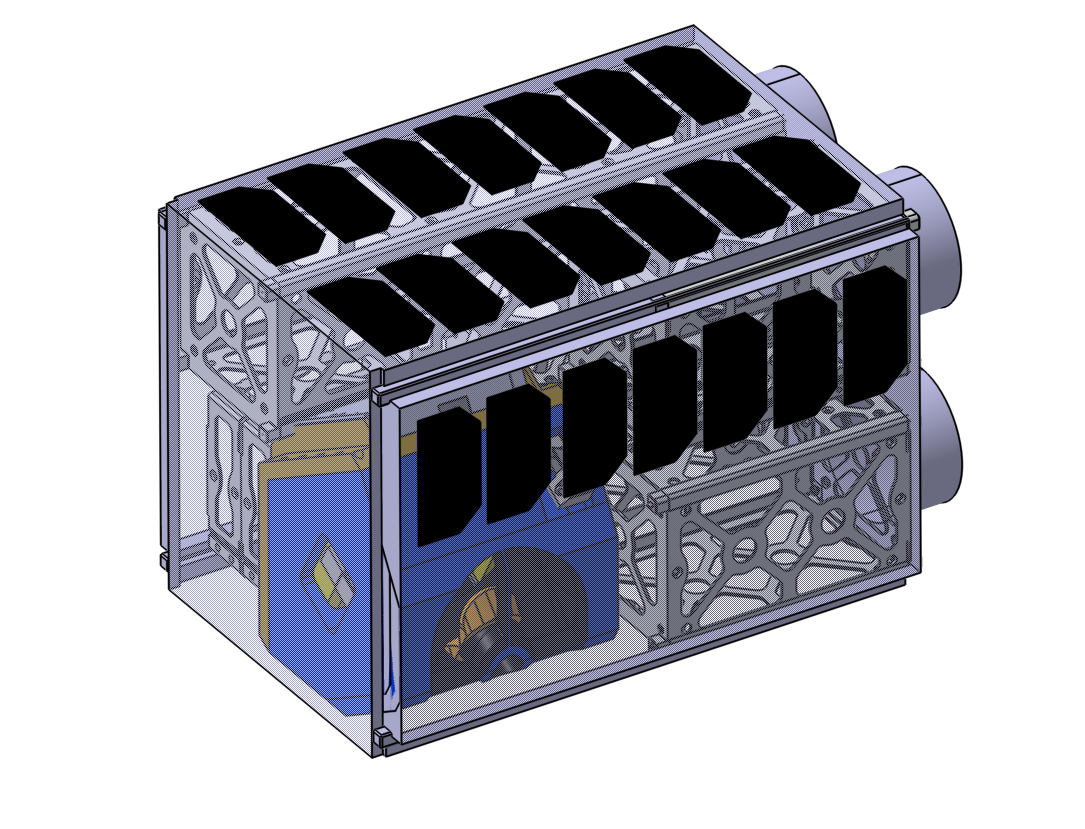}
\caption{ The telescope design (blue part) inside a 12U CubeSat. The preliminary design supposes that 3U (2x1.5U) are used for the propulsion (GTO to GEO), 1.5U for the avionics, 1.5U for attitude control, and 1U for the payload electronics.
\label{fig:riri}}
\end{figure}

\begin{enumerate}
\item a low power laser ($\leq 200\,$mW) at telecom wavelength ($1.55\,\mu$m)
\item fibered optics only, no bulk optics apart from the primary and secondary mirrors 
\item free movement of the spacecraft (ballistic trajectory), no thrusters during science operation
\item because of 3., no free falling test mass, but post processing correction of external forces
\item volume limited to a 12-U CubeSat (Figure~\ref{fig:riri})
\item because of 5., no sun-radiation shielding
\end{enumerate}

The main change, with respect to a simple scaling down of a LISA-type mission, is the absence of a Gravitation Reference Sensor (GRS). It means that the spacecraft must, by itself, act as a test mass. Post processing of the displacements of its center of mass will therefore be paramount.

In section~\ref{sec:2} we will present the concept of the interferometer. In section~\ref{sec:3} the problems caused by the absence of a test mass. Section~\ref{sec:4} will present the optical payload, and section~\ref{sec:5} concludes. 

\section{Principle of the SAGE interferometer}
\label{sec:2}

\begin{figure}
\centering
\includegraphics[scale=0.4]{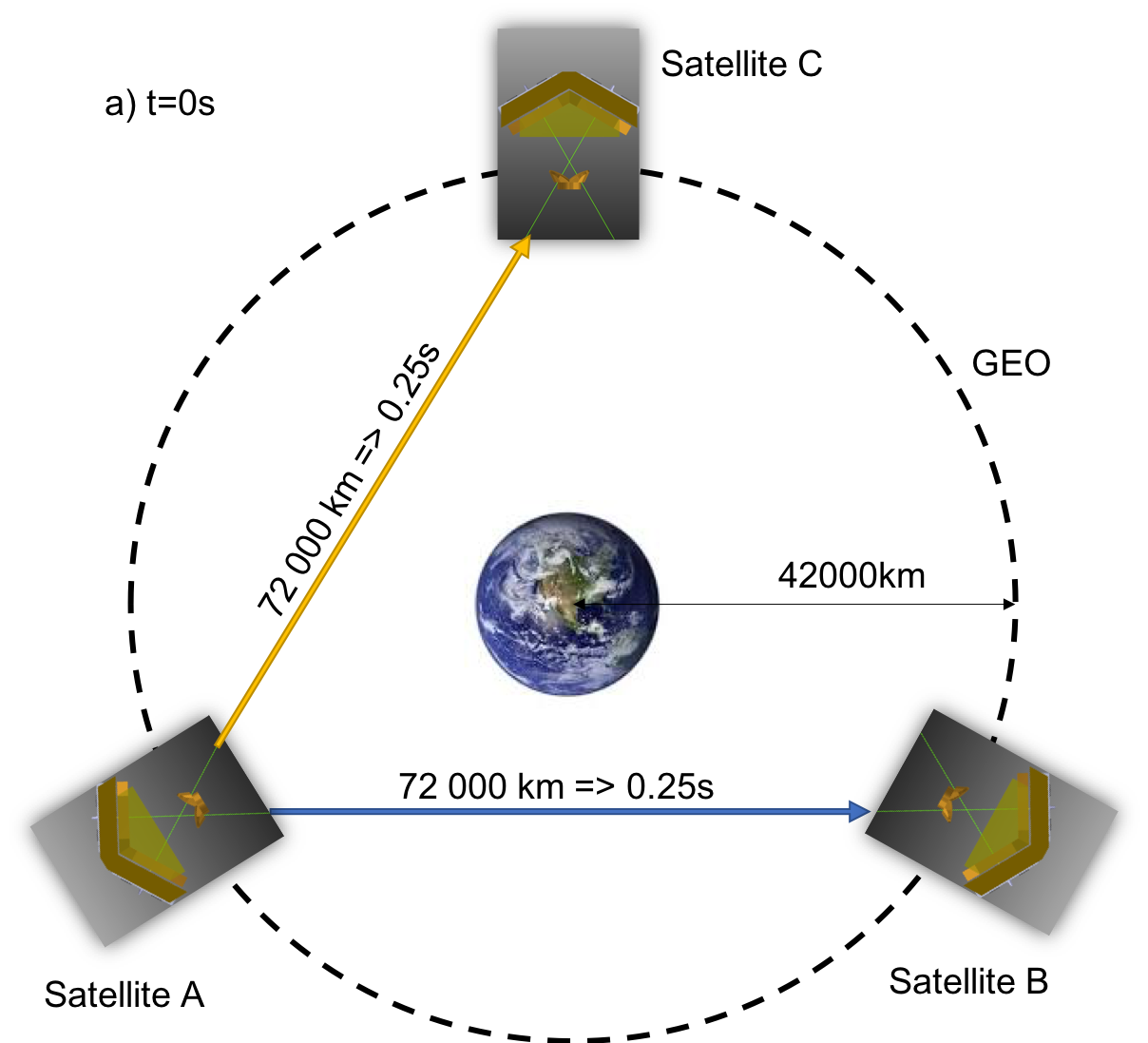}
\includegraphics[scale=0.4]{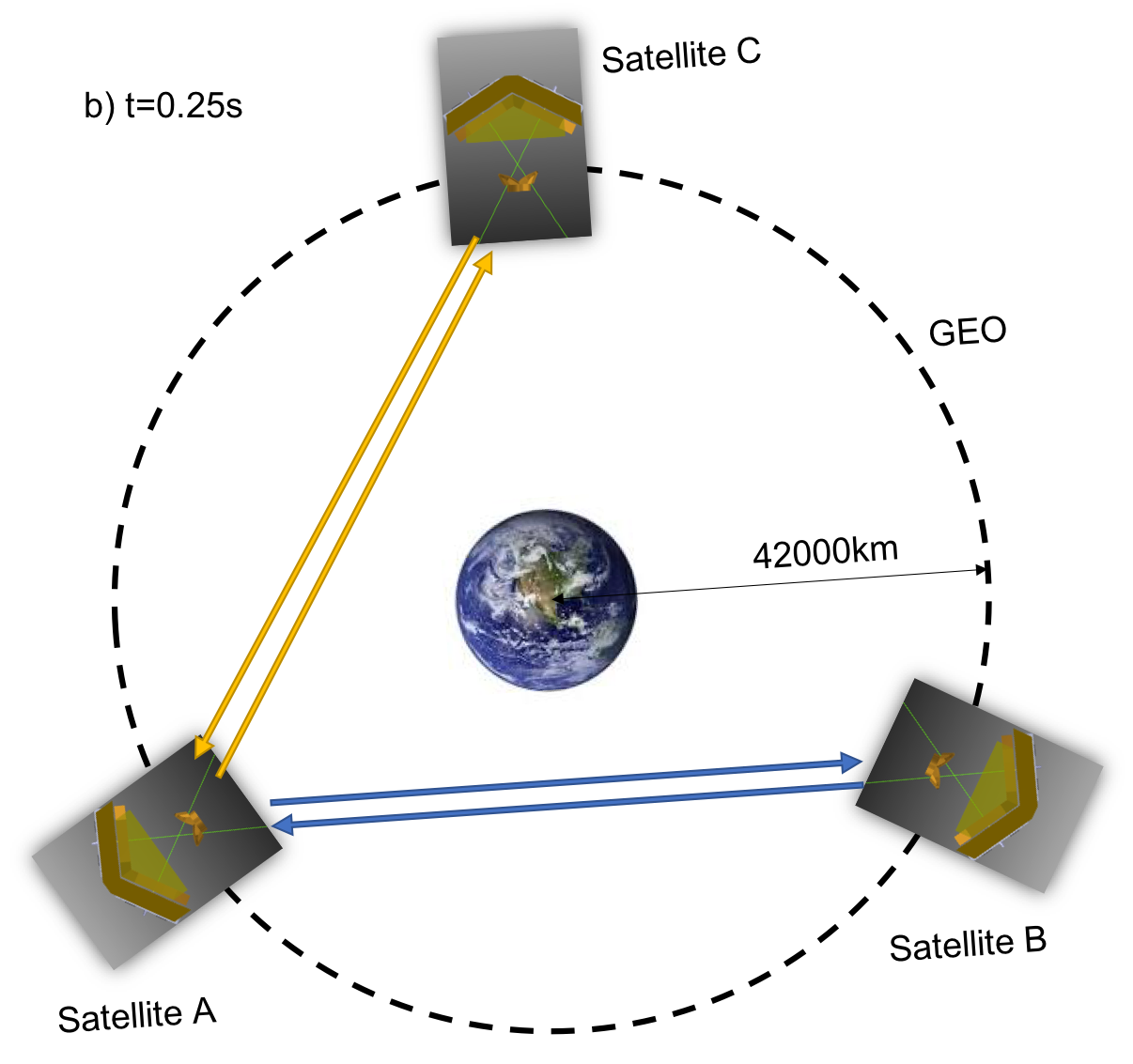}
\includegraphics[scale=0.4]{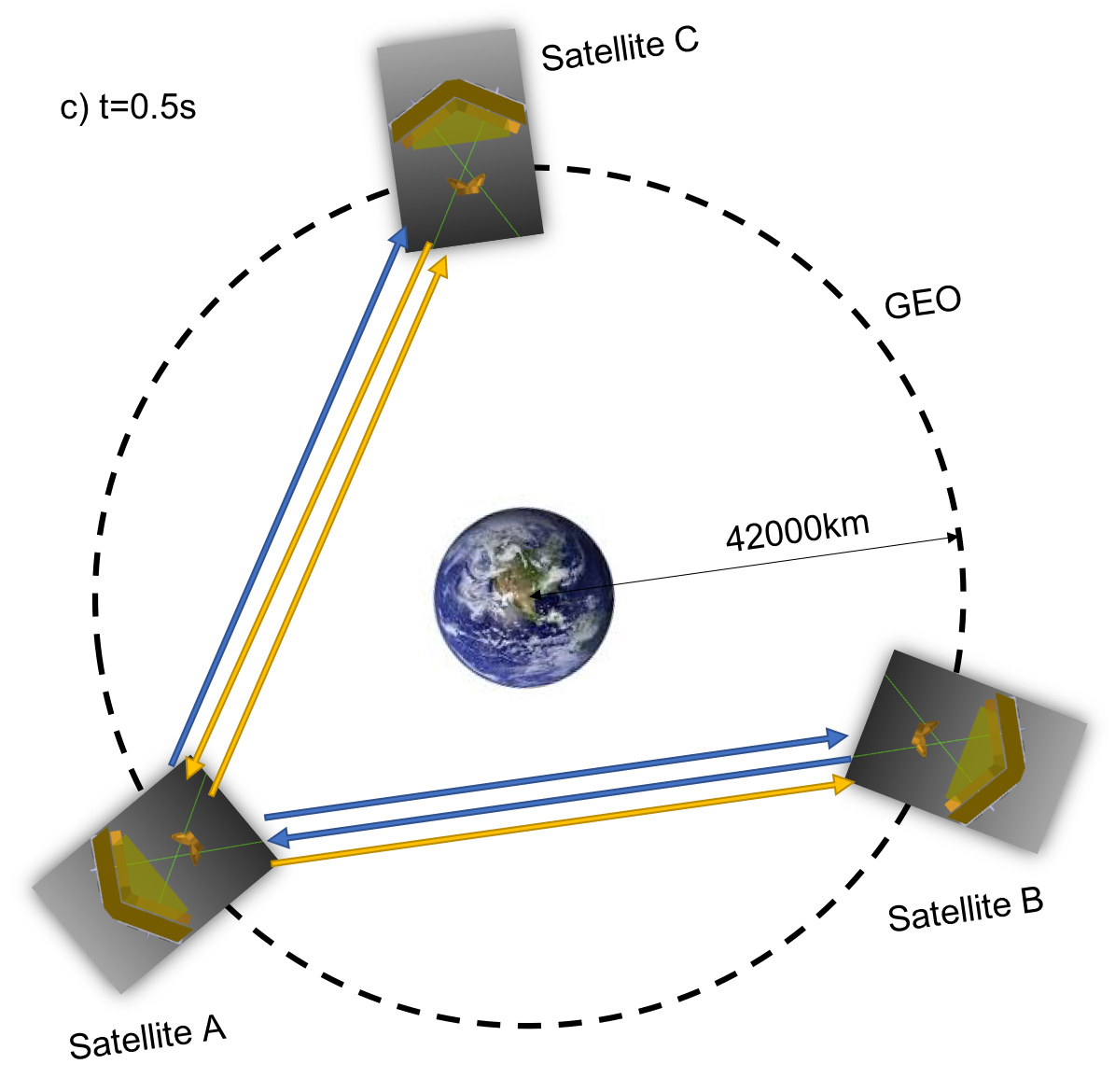}
\includegraphics[scale=0.4]{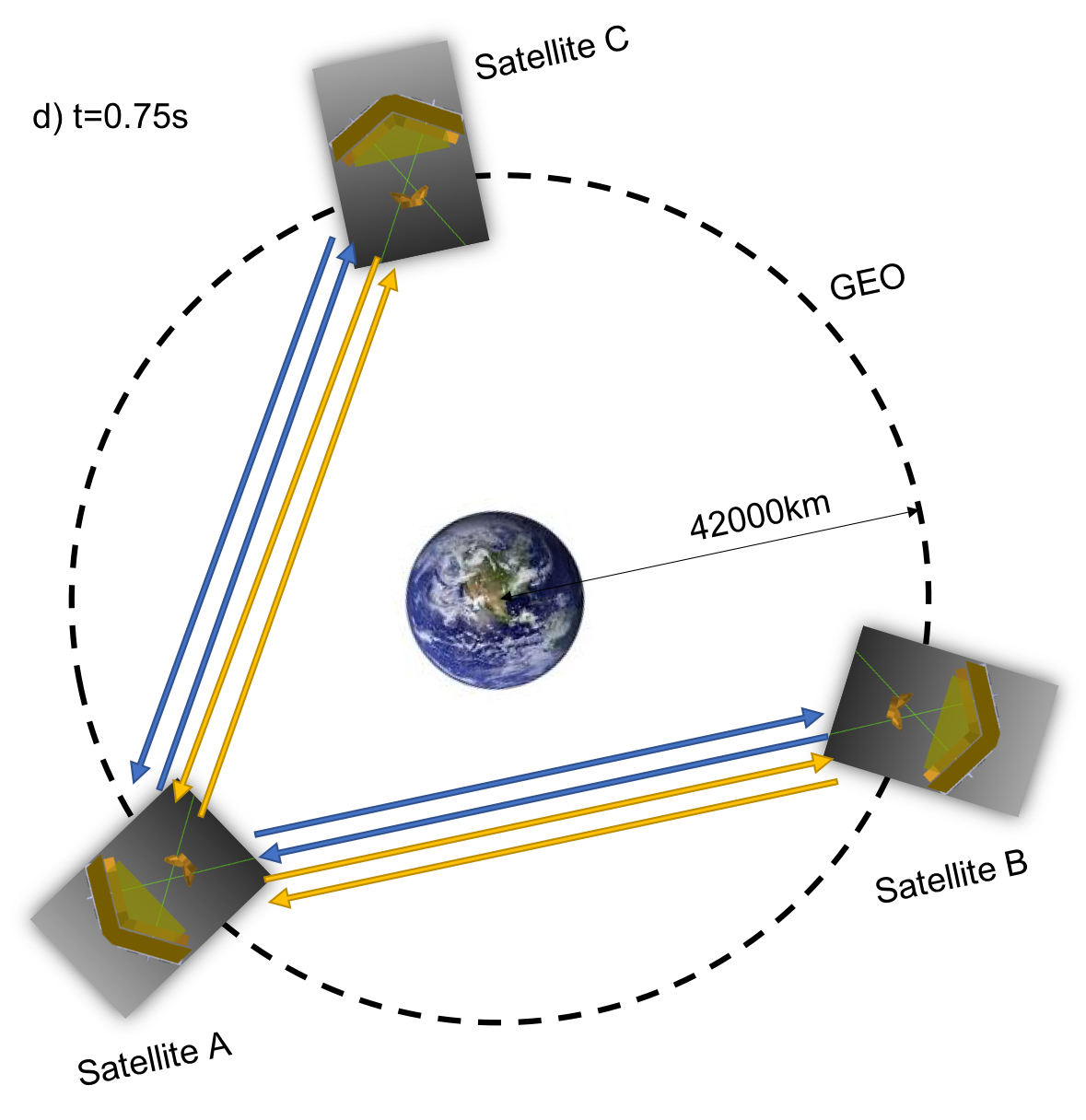}
\caption{Sagnac configuration for a two arms interferometer. The set-up produces the TDI measurement described in Eq.~(\ref{eq:TDI}). {\bf a)} At $t=0$, a beam is launched from satellite A toward satellites B and C. {\bf b)}  At $t=0.25\,$s, the beams are reflected back toward  spacecraft A. {\bf c)}  At $t=0.5\,$s, the beam that returned from spacecraft B is reflected toward C, while the beam from spacecraft C is reflected toward B. {\bf d)} At  $t=1\,s$, the two beams are reflected back to satellite A. All beams have traveled the same distance, and the final measurement can be done with an optical path difference close to 0 even if, during the process, the triangle has rotated by $360^\circ\cdot$1s/1day. Note that the same measurement is done simultaneously on each spacecraft (not just centered on A).
\label{fig:constel}}
\end{figure}

\begin{figure}
\centering
\includegraphics[width=0.75\textwidth]{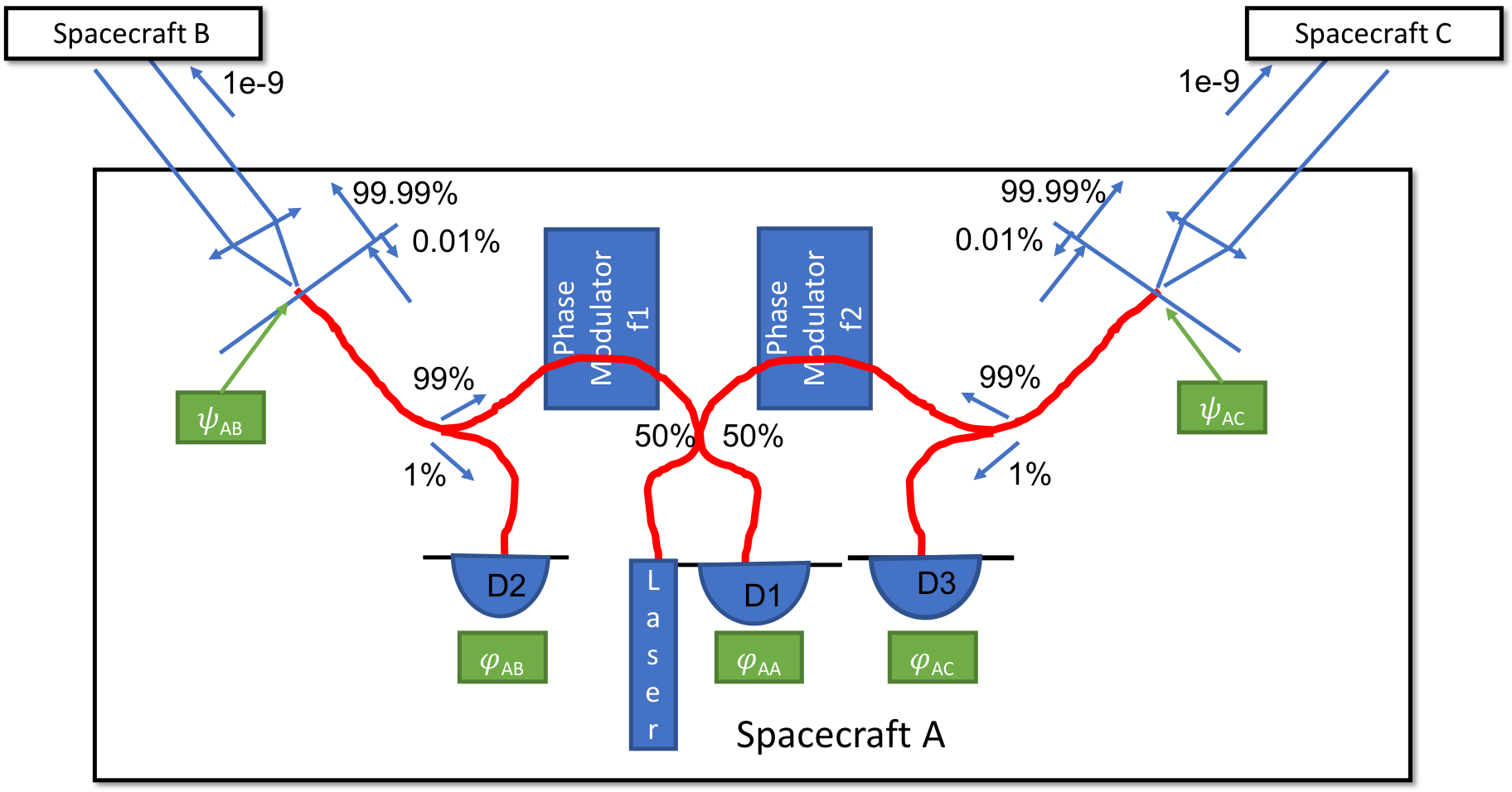}
\caption{ Representation of the optical setup for one of the spacecrafts. The red curves are single mode fibers. One laser beam at $1.55\,\mu$m is split into two, phase modulated thanks to two LiNbO3 phase modulators, and collimated in direction of the two other spacecrafts. The two beams are also weakly back-reflected into the fibers from the fiber extremity (where the beam leaves the fiber). At this point, the light also interferes with the incoming laser beams from spacecrafts 2 and 3. Diodes 2 and 3 measure the phase difference between the two outgoing beams and the two incoming beams. Last, diode D1 measures the internal phase difference between the two outgoing beams.
\label{fig:schema}}
\end{figure}

The detector is based on 3 spacecrafts in geostationary orbit (GEO). Or more exactly slightly above the GEO orbit, in the so-called graveyard orbit \cite{2005ESASP.587..373J}. Each spacecraft is at an altitude of around 36\,000\,km, with the full interferometer forming an equilateral triangle of side 73\,000\,km (Figure~\ref{fig:constel}).

The interferometer will use time delayed heterodyne interferometry (TDI) \cite{2014LRR....17....6T} at $1.55\,\mu$m. Each spacecraft is equipped with a single 200\,mW fibered ECL laser. After dividing the beam thanks to a 50/50 fiber splitter, the two telescopes send two beams at 60 degrees from each other to the other two spacecrafts. For each spacecraft $i$, there are two reference optical positions: $\psi_{ij}$ and $\psi_{ik}$. These positions correspond to the two semi-reflective extremities of the single mode fibers (see Figure~\ref{fig:schema} for a representation of the optical layout inside the satellite). They are used as reference position because they are the last flat optical surfaces before collimation towards, respectively, spacecrafts $j$ and $k$. They are purposely used to monitor the non-common path in the optical setup between the two external metrology measurements (diode $D2$ and $D3$ in figure~\ref{fig:schema}). This monitoring is made thanks to the diode $D1$.

On board each spacecraft, three phases corresponding to three optical path differences (OPD) are measured (denoted $\phi_{ij}$). As mentioned in the previous paragraph, one OPD is measured on diode $D1$ between the two reference positions inside the satellite ($\phi_{ii}=2(\psi_{ij}-\psi_{ik})$). Two other OPDs ($\phi_{ji}$) are measured between the position of the outgoing fiber $\psi_{ij}$ and the corresponding fiber on another satellite (with a delay equivalent to the time of flight). The TDI measurement is a combination of all the optical path measurements such that the absolute optical phase values $\psi_{ij}$ become irrelevant. Here we consider an unequal-arms Michelson configuration, a version of the Sagnac where the photons sweep through the interferometric arms 4 times.
The below equations indicate the relations between OPD and optical phases at the reference positions, for satellite A. The optical length delay between two satellites are denoted $L_{ij}$ for a photon going from spacecraft $i$ to $j$. Note that because the interferometer is in a geostationary orbit, the equilateral triangle rotates at a speed of 360\,deg each days. This orbital period results in a difference between  $L_{ij}$ and $L_{ji}$ of almost a kilometer. Using the terms $\Delta_{ij}=L_{ij}/c$:
\begin{eqnarray}
\phi_{AB}(t)&=&\psi_{AB}(t)-\psi_{BA}(t-\Delta_{BA})-L_{BA}(t) \label{eq:1}\\
\phi_{BA}(t)&=&\psi_{BA}(t)-\psi_{AB}(t-\Delta_{AB})-L_{AB}(t) \\
\phi_{AA}(t)&=&2(\psi_{AC}(t)-\psi_{AB}(t))\\
\phi_{AC}(t)&=&\psi_{AC}(t)-\psi_{CA}(t-\Delta_{CA})-L_{CA}(t) \\
\phi_{CA}(t)&=&\psi_{CA}(t)-\psi_{AC}(t-\Delta_{AC})-L_{AC}(t) \label{eq:5}
\end{eqnarray}

The TDI measurement is a linear combination of the 5 measurements above. It is a Sagnac configuration in the sense that the measurement follows two beams in opposite directions:
\begin{equation}
\begin{split}
h(t)&=\phi_{AB}(t)+\phi_{BA}(t-\Delta_{BA})+\phi_{AC}(t-\Delta_{BA}-\Delta_{AB})+\phi_{CA}(t-\Delta_{BA}-\Delta_{AB}-\Delta_{CA})\\
&\ -\phi_{AC}(t)-\phi_{CA}(t-\Delta_{CA})-\phi_{AB}(t-\Delta_{CA}-\Delta_{AC})-\phi_{BA}(t-\Delta_{CA}-\Delta_{AC}-\Delta_{BA})\\
&\ +\phi_{AA}(t)/2-\phi_{AA}(t-\Delta_{BA}-\Delta_{AB})/2+\phi_{AA}(t-\Delta_{CA}-\Delta_{AC})/2\\
&\ -\phi_{AA}(t-\Delta_{BA}-\Delta_{BA}-\Delta_{CA}-\Delta_{CA})/2
\end{split}
\label{eq:TDI}
\end{equation}
which, using Eqs.\,(\ref{eq:1}) to (\ref{eq:5}), relates to the optical path lengths between the satellites by the following relation:
\begin{equation}
\begin{split}
h(t)=&-L_{BA}(t)-L_{AB}(t-\Delta_{BA})-L_{CA}(t-\Delta_{BA}-\Delta_{AB})-L_{AC}(t-\Delta_{BA}-\Delta_{AB}-\Delta_{CA})\\
&+L_{CA}(t)+L_{AC}(t-\Delta_{CA})+L_{BA}(t-\Delta_{CA}-\Delta_{AC})+L_{BA}(t-\Delta_{CA}-\Delta_{AC}-\Delta_{BA}) 
\end{split}
\label{eq:eq_h2}
\end{equation}

In Eq~(\ref{eq:eq_h2}), all the phase values $\psi_{ij}$ have been canceled out. It means that the absolute phase of the laser beam does not matter, which is an important property of the Sagnac combination.

Another important advantage of the Sagnac configuration is that  $h(t)\approx 0$ if the interferometer formation is constant ($L_{ij}(t)=cst$). A concrete example is the case of a satellite drifting on his orbit. The satellite average speed is of the order of $3\,$km/s. For a differential speed of the order of 1\,m/s \cite{2015CQGra..32r5017T}, then the terms $L_{ij}(t)-L_{ij}(t-0.75s)$ are smaller than one meter, well within the coherence length of the laser. This resolve one important problem of optical interferometry from space: the coherence length of the laser. Any measurement far away from the zero optical path would blur the fringes: wavelength instabilities translate into an optical path error. This phenomenon is also called the "phase noise". In the case of a geostationary constellation, it is especially relevant due to the speed of the satellites.

\section{External forces}
\label{sec:3}

Because SAGE abandons free fall, no reference mass can be kept internally in a free falling state. We are thus loosing a precise reference position. Instead, the extremities of the fibers are the reference positions, and the whole satellite is the equivalent of the test mass. Any force from outside will affect the center of mass of the satellite, and hence, the OPD measurement.

The most important environmental forces in space are caused by solar radiation pressure and solar wind. Both noise terms have two components. A red component, which corresponds to variations in the particle density. The second component is a white noise caused  by the shot noise of the particles.

But this is not all. The position of the barycenter of the satellite can also move due to thermal effects. This too has to be accounted for, especially in the context of expansion and deformation of the platform.

\subsection{Solar Radiation Pressure}

\begin{figure}
\centering
\includegraphics[width=0.8\textwidth]{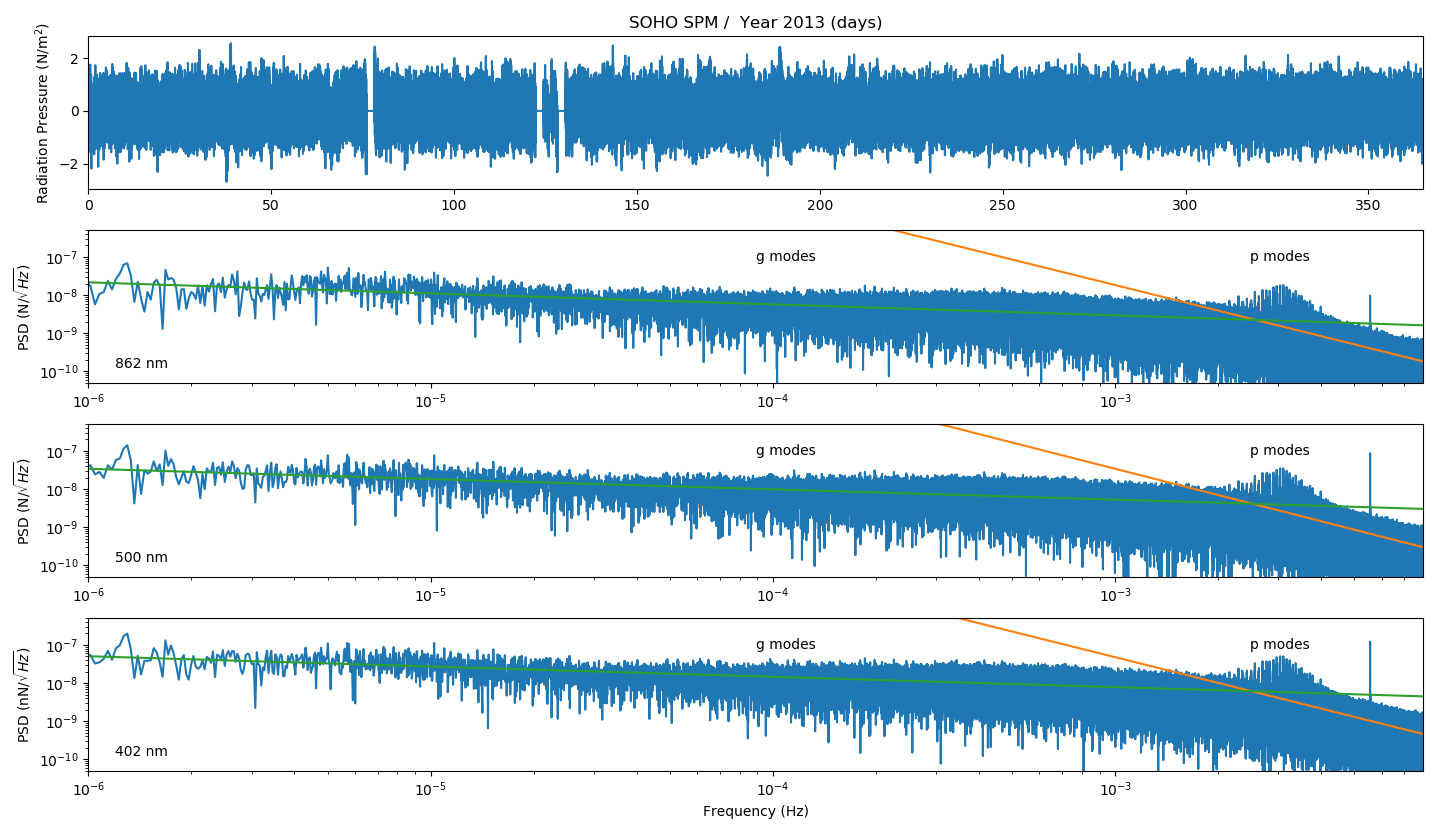}
\caption{ {\bf Solar Photons:} top panel, the flux variation observed from the SOHO satellite in the 0.1 to 50\,nm band, scaled to the average solar radiation pressure at 1\,AU which is $9.08\,\mu$N/m$^2$. Lower panels: power spectrum density (blue curve), of the three color channel of the SPM instrument (from top to bottom, the red, green, and blue channels). The green curves is the fit of the gravitational modes, and the orange curves are the fit of the granulation. The 3\,mHz forest lines ($p$-modes) are excluded from the fit.
\label{fig:photons}}
\end{figure}

At 1\,AU from the Sun the solar radiation averages to $9.08\,\mu$N/m$^2$. This acceleration is of the order $10^4$ higher than what is expected from the GW. However, most of the energy is at low frequency: the total solar irradiance (TSI) changes slowly with time. To characterize its power spectral density (PSD), we used data from the VIRGO (Variability of solar IRradiance and Gravity Oscillations) instrument on-board the SOHO observatory. VIRGO contains three channel sunphotometers (SPM) that monitor the irradiance at 402, 500, and 863\,nm with a bandwidth of 5\,nm.\cite{Frohlich95,Frohlich97,2002SoPh..209..247J}.

In figure~\ref{fig:photons} we present a dataset that covers the full year 2013. Data sampling is 60\,s, which means that the maximum frequency (Shannon's frequency) is 8\,mHz. We plotted the three PSDs for the 3 channels. One can clearly see the pressure modes that dominate the PSD around 3\,mHz (resonance of the radiative structure). Below 1mHz, the spectrum is dominated by the gravitational modes (convective structure). The trend, below and after the $p$-modes, are caused by the granulation. This is a red noise which can be fitted by a logarithmic relation: $\sqrt{S_{\rm photon}(f)}=\alpha \times f^\mu$. For respectively the blue, green, and red channels we have obtained: $\alpha = 
9.2\times10^{-15}$, $4.7\times10^{-15}$ and $3.9\times10^{-15} N/$m$^2/\sqrt{\rm Hz}$ and $\mu = -2.22, -2.28$ and $-2.23$. The difference in amplitude between the three channels is a consequence of the optical depth of the Sun as a function of the wavelength.

Weighting the PSDs  by the solar 6000\,K black body yields to the relation:
\begin{equation}
S_{\rm photon \ red}(f)=4.8\times10^{-15} f^{-2.25} N/m^2/\sqrt{\rm Hz}
\end{equation}
At the same time, the shot noise can be deduced from the ratio between total energy ($P_{\rm photon}=1000 \,$Wm$^2$) and the energy of a $600$\,nm photon:
\begin{equation}
S_{\rm photon \ shot}(f)= 9.08\times10^{-6} \times \sqrt{h\nu/P_{\rm photon}}\approx 10^{-16} N/m^2/\sqrt{\rm Hz}
\end{equation}
of amplitude well below the irradiation variation.

\subsection{Solar Wind}

\begin{figure}
\centering
\includegraphics[width=0.9\textwidth]{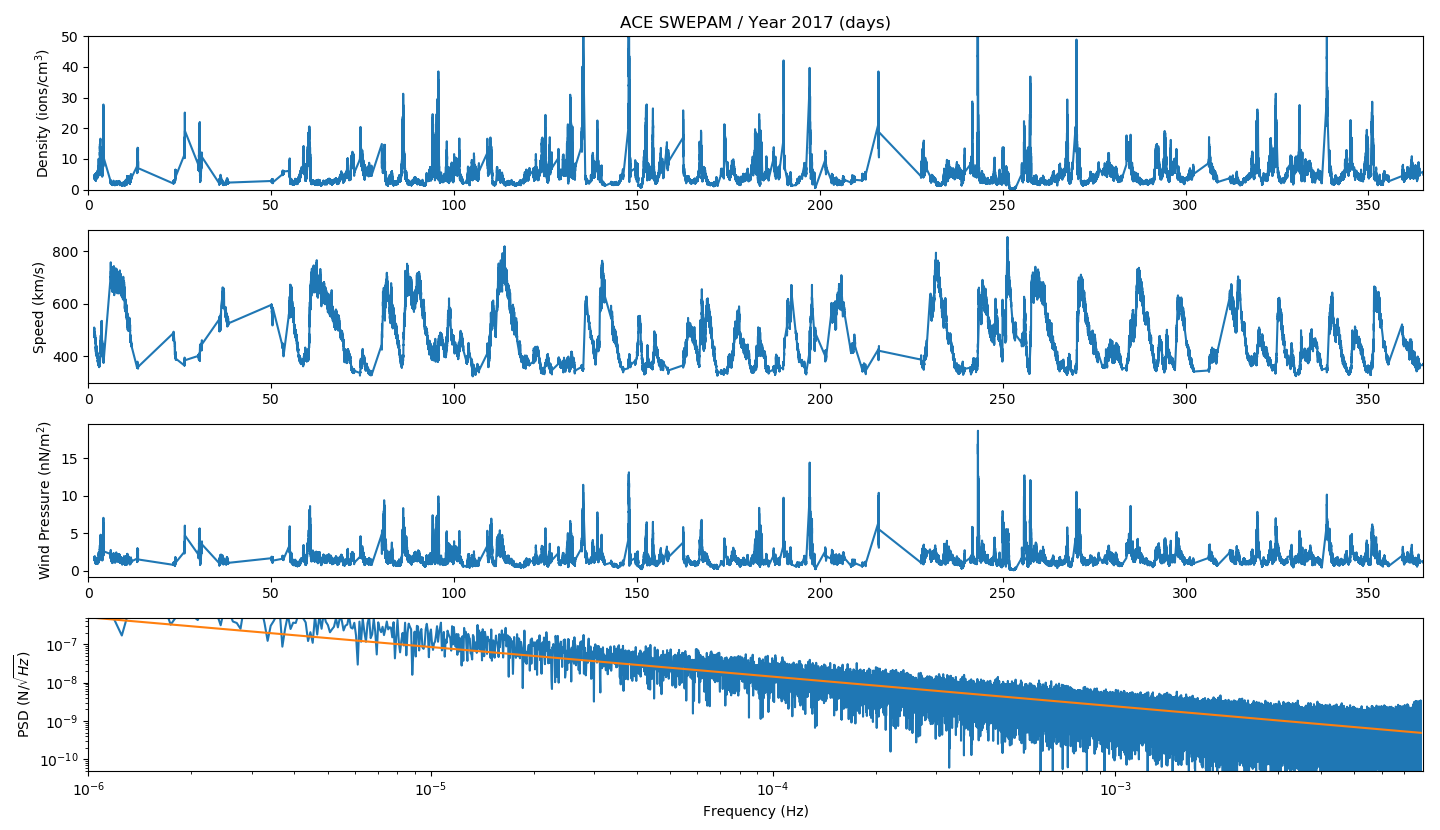}
\caption{ 
{\bf Solar Protons:} top panels: density and speed of solar protons observed during the period 2017 from the SWEPAM instrument aboard the ACE observatory. The observatory is situated close to the L1 Lagrangian point, and gives us only an estimation of wind variation as the spacecraft will evolve on GEO orbit. Middle lower panel: the resulting pressure from the protons on the spacecraft. Lower panel: power spectral density of the solar wind pressure.
\label{fig:protons}}
\end{figure}

To estimate the amount of solar protons that will be accelerating the satellite, we restricted ourselves to the low speed ones ($v<<c$) which are emitted by the Sun as the solar wind. This wind is monitored from the L1 Lagrange point by the SWEPAM instrument on board the ACE satellite. The instrument monitor both the speed $v$ and the density $d$ of the ions in the solar wind. The density is typically of the order a several particles per $cm^3$. However, it is especially sensitive to interplanetary shocks, which cause a sudden rise in density of particles (up to $100\,$cm$^3$), followed by an acceleration of the wind. These spikes in the solar wind momentum dominate the power spectrum.

Figure~\ref{fig:protons} plots one year of data of ion density and speed. The force applied to the satellite, in N/m$^2$, is calculated from the mass of a proton $m_p$: $F_{\rm protons}=m_q*d*v^2$. The fit of the power density gives :
\begin{equation}
S_{\rm proton \ red}(f)=1.1\times10^{-11} f^{-0.744} N/m^2/\sqrt{\rm Hz}
\end{equation}
while the shot noise can be derived from the average density ($\approx 5$ per cm$^3$) and speed ($\approx 465$\,km/s) of protons :
\begin{equation}
S_{\rm proton \ shot}(f)= 3\times10^{-10} \times \sqrt{2\times10^{-12}}\approx 2\times 10^{-16} N/m^2/\sqrt{\rm Hz}
\end{equation}

\subsection{Temperature sensibility}

The reference positions are  the extremities of the fibers ($\psi_{ij}$ in figure~\ref{fig:schema}). Even if the center of mass is stable, the position of the two reference positions may move with respect to it. The simplest example is a rotation of the satellite. This rotation is fortunately not an issue because it can be measured and post-processed thanks to our knowledge of the optical orientation with respect to the two other spacecrafts.

However, a major problem will come for the thermal modification of the spacecraft, even if the full CubeSat is made of a material with a very low thermal expansion coefficient (eg, SiC). For example, with a material of thermal expansion of $2\times 10^{-6}$\,K$^{-1}$, and considering the typical size of 10\,cm, it would creates a noise of the order of 200\,nm/K$^{-1}$.

Engineering a design that makes the center of mass unaffected by temperature gradients will be a major challenge. In the worst case, it requires a monitoring of the spacecraft temperatures at a level of $10^{-4}$\,K$/\sqrt{Hz}$, which means that the effect on the reference positions will be maintained below 20\,pm$/\sqrt{Hz}$.

\section{Opto-mechanical design}
\label{sec:4}

\subsection{The telescopes}

\begin{figure}
\centering
\includegraphics[width=0.45\textwidth]{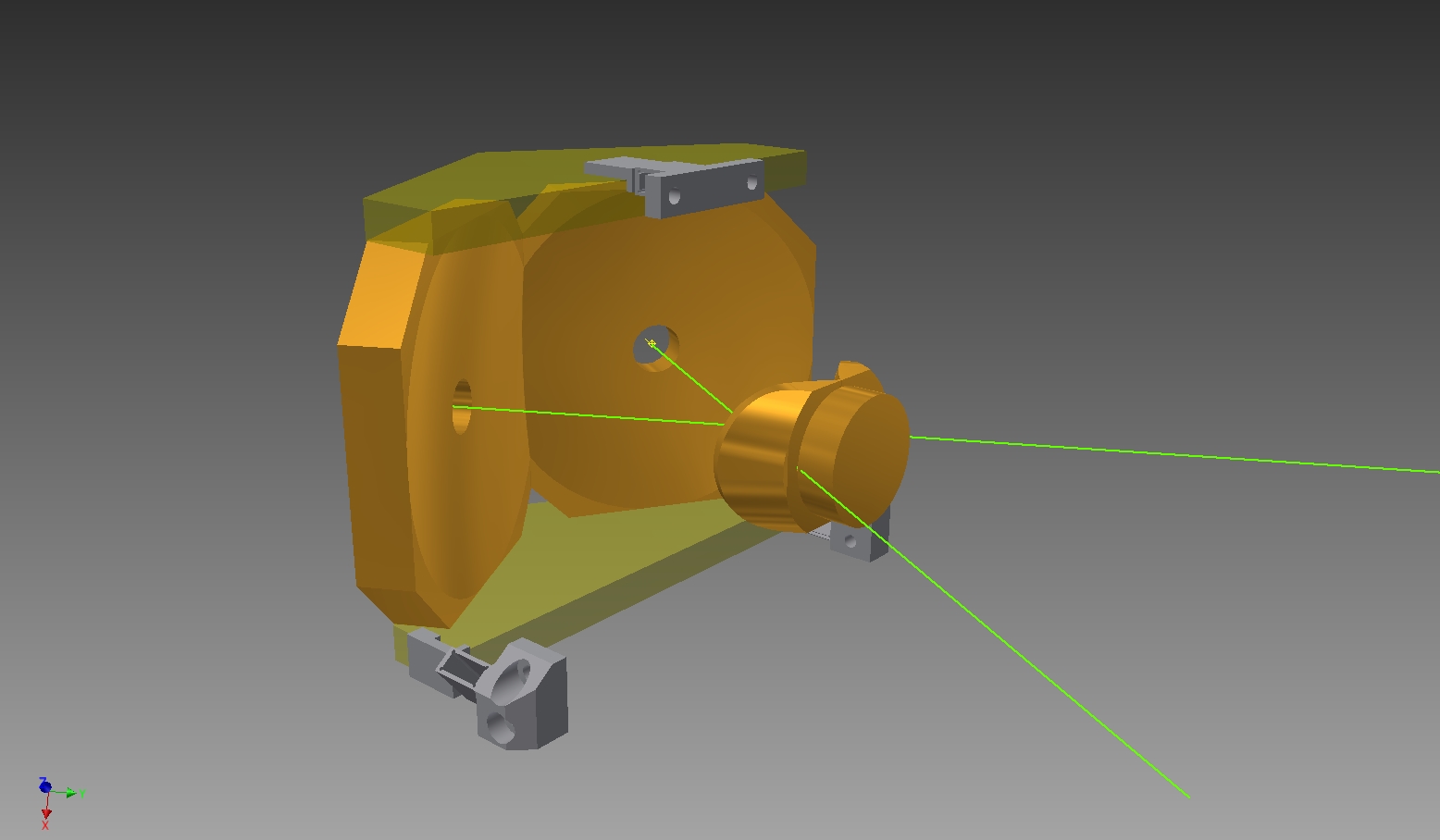}
\includegraphics[width=0.45\textwidth]{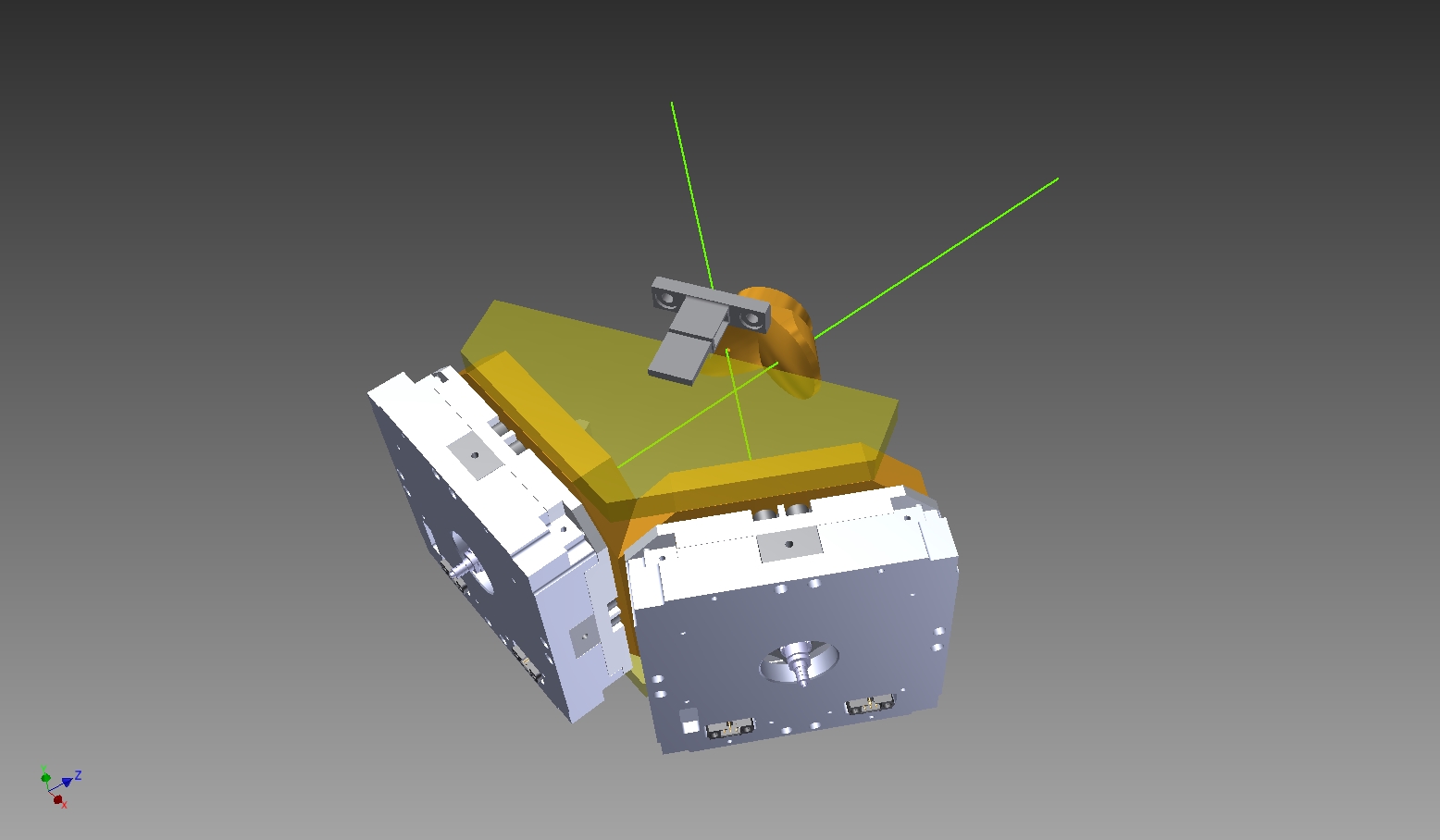}
\includegraphics[width=0.45\textwidth]{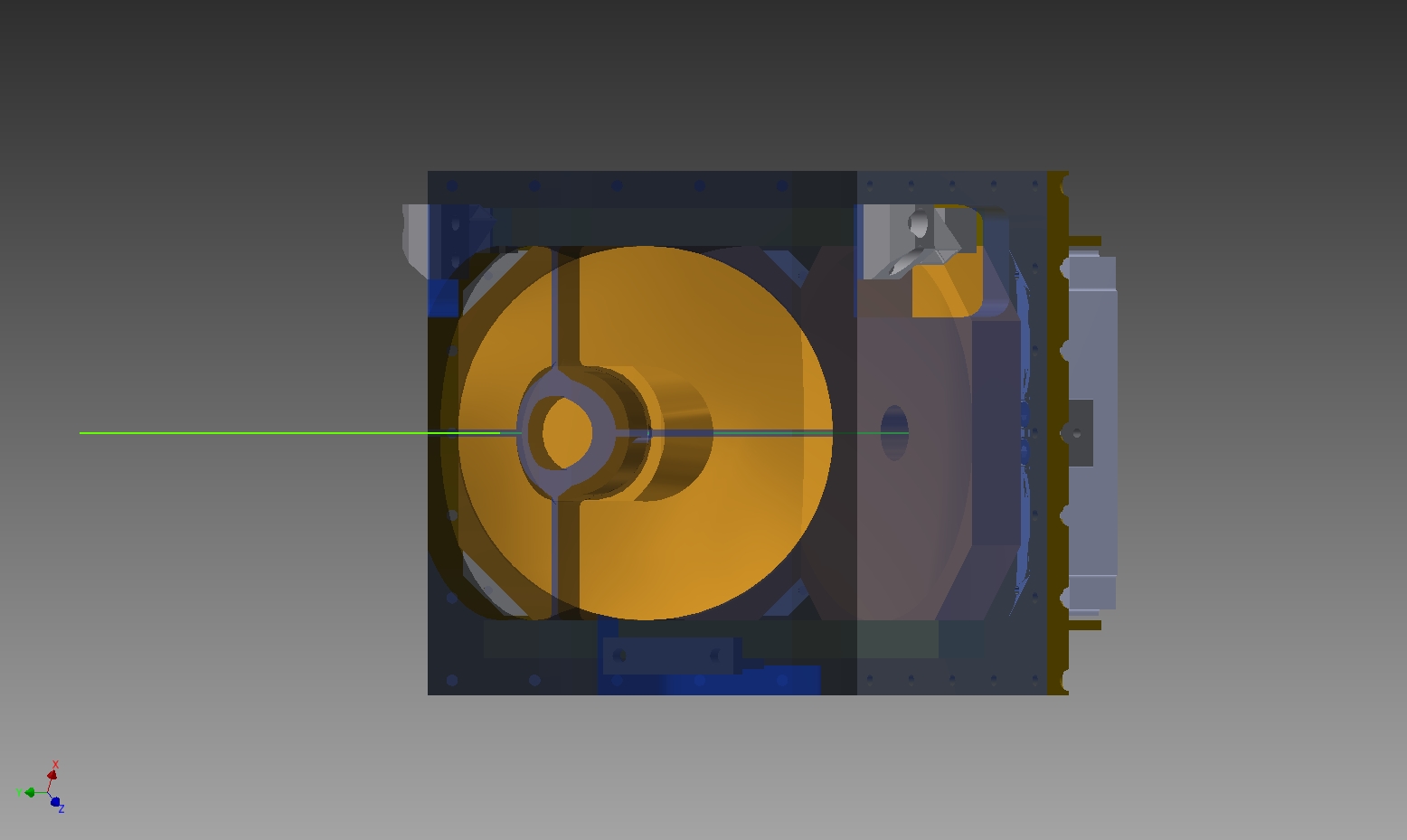}
\includegraphics[width=0.45\textwidth]{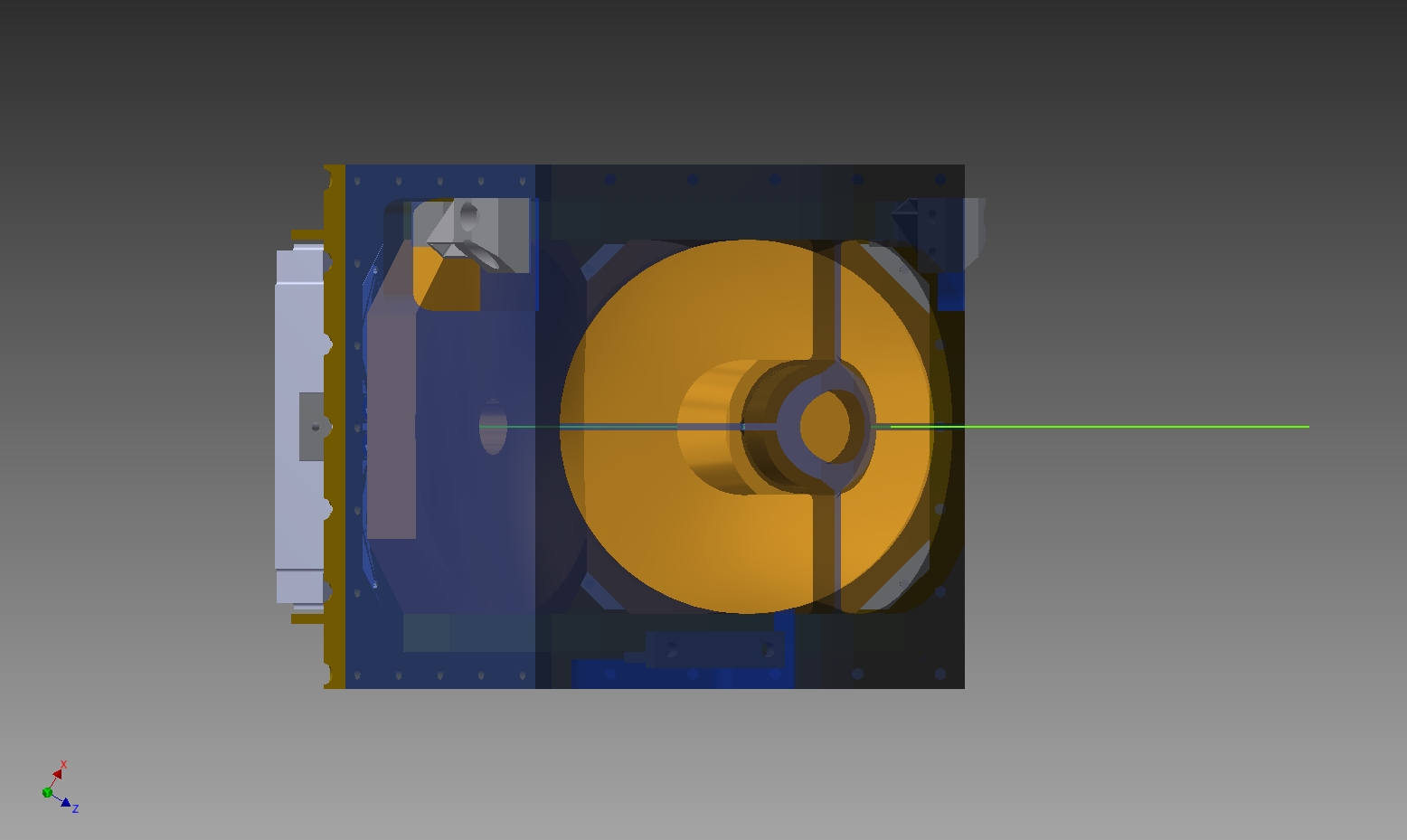}
\caption{3D rendering of the optical part of the spacecraft. The two telescopes are intertwined with a 60 degrees orientation shift between the two. The Zerodur primary and secondary mirrors are maintained by molecular cohesion. Fine pointing is obtained through a two axes piezo stage (in gray in the upper right panel). The structure of the telescopes, in Invar material, is in blue in the lower panels.\label{fig:opto}}
\end{figure}

Each satellite will manoeuvre to maintain the equilateral formation. Thus each satellite must be launching two beams at $60^\circ$ separation. For compactness, but most importantly, for robustness, the two telescopes are intertwined. The two primary mirrors, in Zerodur and of size 10\,cm, are maintained together by molecular cohesion. The two secondary mirrors, also in Zerodur and of diameter 36\,mm, are also similarly glued and maintained together by molecular cohesion. To maximize thermal stability, the two M2s are kept in position with respect to the M1s through a common structure in invar material.

The opto-mechanical concept of the instrument is presented in Figure~\ref{fig:opto}. It is driven by the following constraints:
\begin{itemize}
\item Stability of the $60^\circ$ angle of the two telescopes satellites to project an equilateral triangle in space
\item Thermal environment
\item Available space inside the 12U platform
\end{itemize}
Thermal environment and stable angular reference between the two telescopes of each satellite were the main reasons to choose a Zerodur + invar assembly instead of a single material for the structure and the optics. For thermal stability of the angular reference and space constraint, on axis optics was preferred at the cost of a central obstruction of 14\% with a 100\,mm pupil diameter. The two M1 and M2 assemblies are glued during the alignment phase to provide the $60^\circ$ reference angle. Molecular cohesion is then applied to the M1 assembly between the upper and lower zerodur interface plates. Both M1 and M2 units are then integrated and glued to the invar monobloc body  of the telescope. The invariant point of the differential thermal expansion of the complete assembly is as close as possible to the M2 unit. An accurate temperature sensing all over the two telescopes assembly provides input to the model of the differential thermal expansion (Invar/Zerodur) on the $60^\circ$ reference pointing. This bias is taken into account in the correction and error budget of the full interferometer metrology. 

The beams are launched from a single mode fiber at the focal point of each telescope. Fine pointing is achieved thanks to a piezo stage with range 400\,$\mu$m. This range, combined to the f/D value of 0.25, gives a field of view of $0.057^\circ$. This means that during operation, the position of the satellites must be accurate to a projected distance of 73\,km ($73\,000\times0.57^o$).

\subsection{Collimating and coupling the single mode light}

\begin{figure}
\centering
\includegraphics[width=0.85\textwidth]{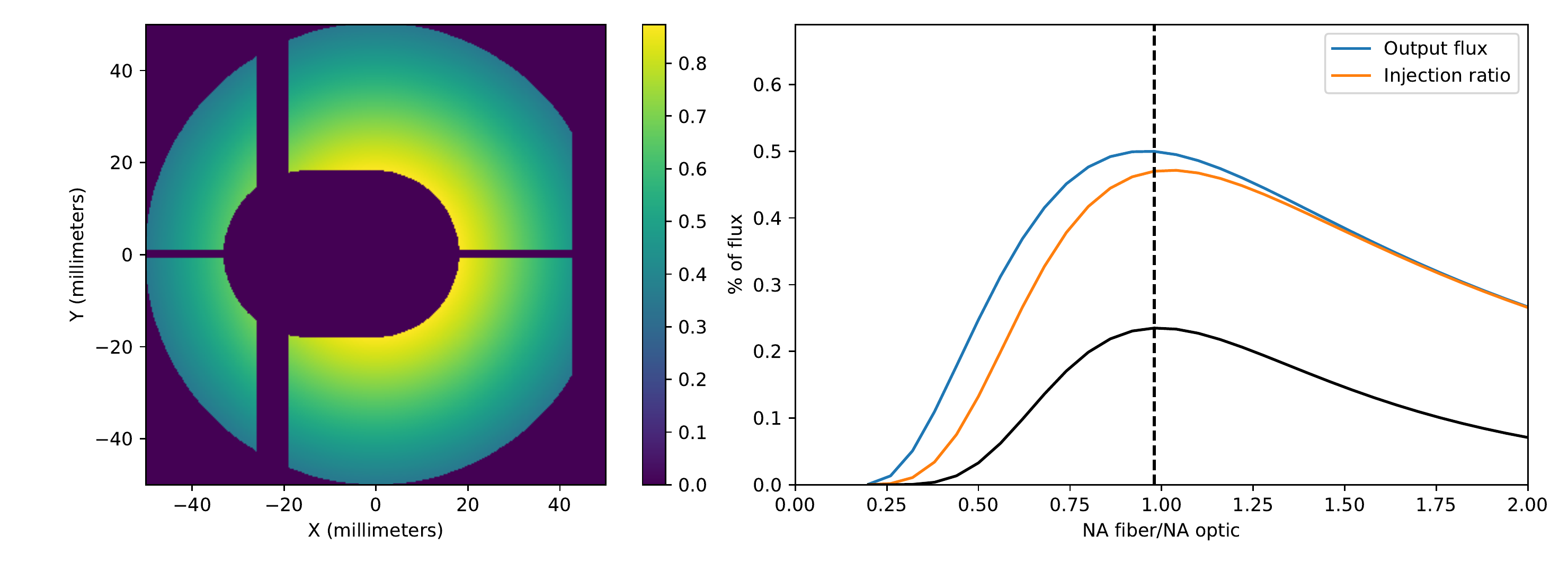}
\caption{ Left : The pupil $P(x,y)$ including the central obstruction and the shadow caused by the spider arms. The energy of the output beam ($|E(r)|^2$) is plotted on the color scale. Right: the blue curve traces the throughput as a percentage of the flux emitted by the single mode fiber ($\rho_{\rm Through}$), a function of $w/$NA$_{\rm optics}$. The orange curve is the ratio between the light received in the pupil, and the light injected into the fiber. The black curve is the product between collimation loss and coupling loss.
\label{fig:inject}}
\end{figure}

The power of the source results from a difficult compromise. On one hand photons are important for a precise metrology measurement, so it is necessary to have a powerfull beam. But on the other hand the thermal stability, consumption level, and non-linear effects in the fibers require moderate power.

To help with this compromise, it is important to maximize the throughput. This comes in two flavors: decreasing the collimating loss and increasing the coupling efficiency. Both are constrained by the mode of the fiber and the shape of the telescope aperture. The shadow from the spider arms and dual-secondary mirrors is presented on figure~\ref{fig:inject}. This shadowing is important due to the fact that we wanted to keep the optics on-axis for thermal stability.

We can approximate the mode of the single mode fiber to be a 2 dimensional Gaussian. Then, the mode writes:
\begin{equation}
    E(r)=\frac{\sqrt{2}}{w\sqrt{\pi}}\exp^{(-r^2/w^2)}
\end{equation}
where $r$ is the radial position in the pupil, and $w$ is the numerical aperture as defined by the flux at $\exp^{-2}$. The total throughput of the system is:
\begin{equation}
    \rho_{\rm through}=\iint P(x,y)  |E(\sqrt{x^2+y^2})|^2 dxdy
\end{equation}
where $P(x,y)$  is the pupil geometry: 1 if the light passes through, 0 otherwise. The optimization of $\rho$ as a function of $w$ is shown in Figure~\ref{fig:inject}. It has a maximum throughput flux for $w=0.96\,$NA$_{\rm optics}$. The  throughput is then 49.6\%, meaning that 50.4\% of the laser power will be lost and/or absorbed by the optical system.

Assuming the receiving satellite is sufficiently distant, we can expect the incoming field to be flat and constant over the pupil. Then, the coupling efficiency is:
\begin{equation}
    \rho_{\rm coupling}=\left|\cfrac{\iint P(x,y)  E(\sqrt{x^2+y^2}) dxdy}{\iint P(x,y) dxdy }\right|^2
\end{equation}
In the end, the important value is $\rho_{\rm through}\times \rho_{\rm coupling}$. It reaches a maximum of $23.5\%$. 

\subsection{Diffraction analysis}

\begin{figure}
\centering
\includegraphics[width=0.9\textwidth]{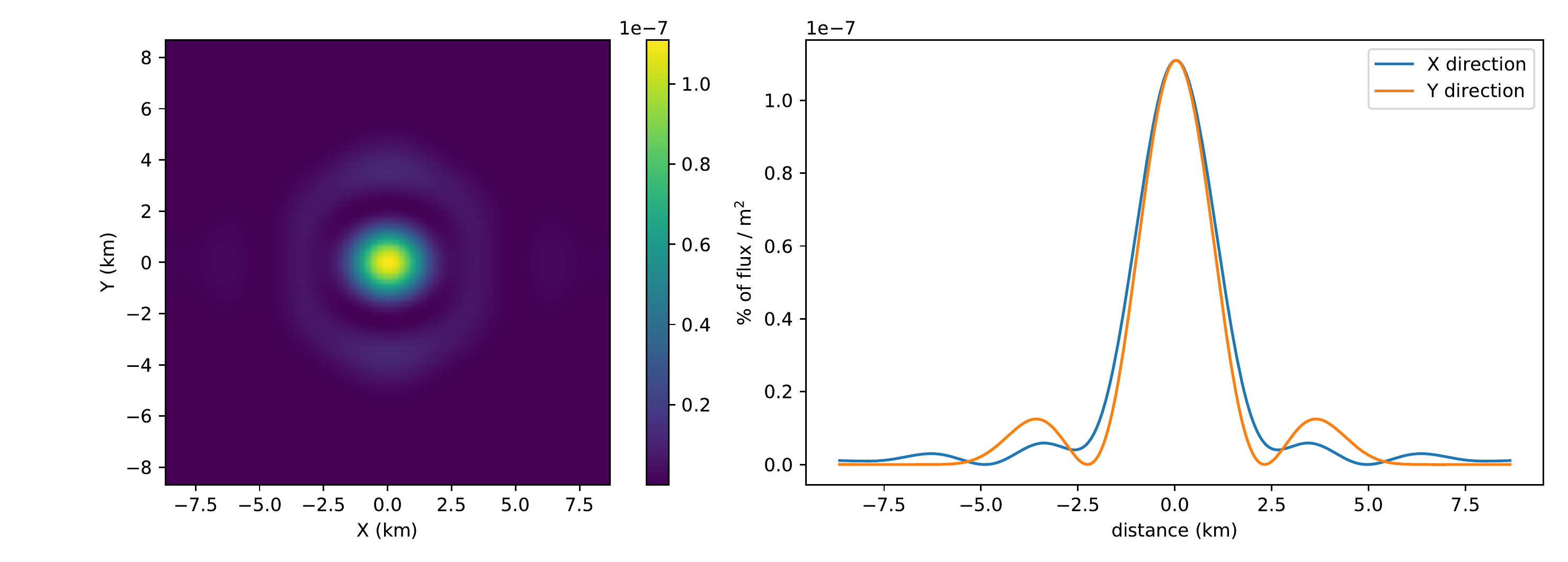}
\caption{ Left: $F(X,Y)$, the diffraction pattern at a distance of 73\,000\,km. Due to the geometry of the pupil the diffraction pattern is not circular (as an Airy pattern would be). The full width at half maximum of the pattern is of the order of 2\,km. Left: cut in $X$ and $Y$ of the diffraction pattern. Centered on the pattern, the maximum flux density is of the order of $1.1\times10^{-7}$ per $m^2$.
\label{fig:diff}}
\end{figure}

The telescope should be large to minimize diffraction, and large to maximize the reception: the received flux increases as the power of 4 with the telescope size.

To calculate the far field flux $F(X,Y)$ at a distance of $Z=73\,000\,$km, we can use the Fraunhofer approximation. The electromagnetic field is the Fourier transform of the field inside the emitting telescope pupil:
\begin{equation}
F(X,Y)=\iint P(x,y) E(x,y) \exp(\frac{-2 i \pi (xX+yY)}{\lambda Z}) dx dy
\end{equation}
where $\lambda=1.55\,\mu$m. The resulting diffraction pattern is presented in the left panel of figure~\ref{fig:diff}. Maximum is achieved on axis where the flux is around $1.1\times10^{-7}$ of the emitted flux per square meter. Integrated over the pupil, we get a throughput due to the diffraction of $\rho_{\rm diffraction}=6.1\times10^{-10}$.

\subsection{Photon error budget}

\begin{figure}
\centering
\includegraphics[width=0.9\textwidth]{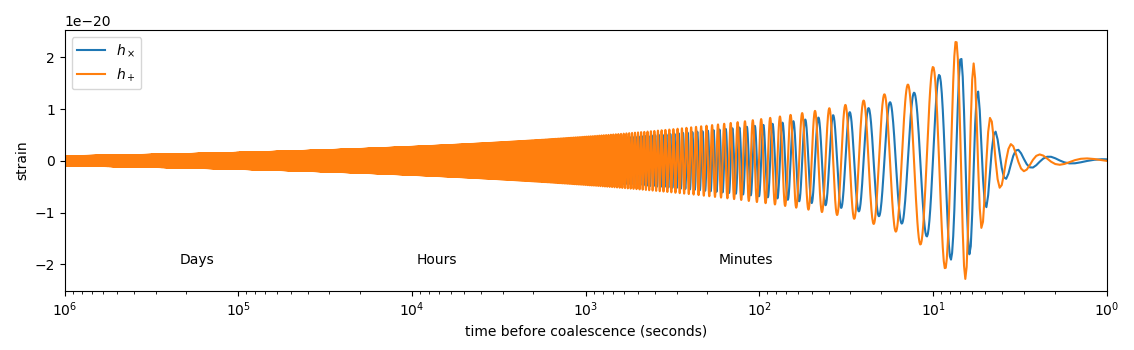}
\caption{ Inspiral amplitude of the two polarized signals for the coalescence of two IMBH of masses $10^4\,M_\odot$, at $z=1$.
\label{fig:inspi}}
\end{figure}

The shot noise is basically the consequence of the quantum limit to measure the state of a photon. It depends on the total number of photons received by the satellite: the more photons we receive, the higher the accuracy on the phase measurement. 
Using a laser power of 200\,mW (divided towards the two other satellites), the photon error depends on the total energy $P_{\rm photons}$ received at the other end. In our case: 
\begin{equation}
P_{\rm photons}=\frac{200\,mW}{2} \rho_{\rm through}\times \rho_{\rm diffraction}  \times \rho_{\rm coupling}  = 15\,pW\,.
\end{equation}
 At wavelength $\lambda=1.55\, \mu$m, the accuracy of the metrology measurement is limited by the photon noise:
 \begin{equation}
 \sqrt{\frac{hc\lambda}{2\pi P_{\rm photons}}}=23\,pm/\sqrt{Hz}
 \end{equation}
 
 The strain of the most energetic gravitational waves is of the order $10^{-21}$. For an interferometer of arm length 73\,000\,km, the effect of the gravitational wave is 0.7\,pm in amplitude. It seems small with respect to $23\,pm/\sqrt{Hz}$. However, the baseline concept is to fit inspiral patterns over several month, even years. In the example case of Fig~\ref{fig:inspi}, the signal can be detected with a signal to noise ratio of 8 in a day of science operation.

\section{Conclusion}
\label{sec:5}

Can we really detect gravitational waves from space without a test mass? We believe so.

On one hand, it is true that a free-falling test mass allows to give a stringent reference point for a drag-free system. This may be the only solution for a high precision measurement. However, a drag-free system complicates the mission drastically: low noise thrust, internal metrology, complicated optical setup etc. 

On the other hand, there is still precious science to be done regardless of the environmental effects. Maybe it is preferable to keep the satellite "quiet", and let it accelerate naturally. This is especially interesting for mid-frequency observations, since the effect of external forces decreases as $1/f^2$. However, it relies on the fact that environmental pressure forces can be measured and subtracted by post-processing.

\acknowledgments 

 S.~Lacour acknowledges support from ERC starting grant No. 639248.
The VIRGO instrument onboard SoHO is a cooperative effort of scientists, engineers, and technicians, to whom we are indebted. SoHO is a project of international collaboration between ESA and NASA. Plot in Figure~\ref{fig:inspi} was generated using the PyCBC software package
\cite{Canton:2014ena,Usman:2015kfa,Nitz:2017svb,alex_nitz_2018_1256897}.

\bibliography{report,sample} 
\bibliographystyle{spiebib} 

\end{document}